\documentclass[journal,twoside,web]{IEEEtran}
\usepackage{tmi}
\usepackage{cite}
\usepackage{amsmath,amssymb,amsfonts}
\usepackage{siunitx}
\usepackage{mathtools}
\usepackage{algorithm}
\usepackage{algpseudocode}
\usepackage[pdftex]{graphicx}
\usepackage{textcomp}
\usepackage{bm}
\usepackage{hyperref}
\def\E{\mathbb{E}}

\def\xbm{{\bm{x}}}
\def\zbm{{\bm{z}}}

\def\zbm{{\bm{z}}}

\def\thetabm{{\bm{\theta }}}

\def\Ncal{{\mathcal{N}}}

\newcommand{\norm}[1]{\left\lVert#1\right\rVert}

\UseRawInputEncoding
\def\BibTeX{{\rm B\kern-.05em{\sc i\kern-.025em b}\kern-.08em
    T\kern-.1667em\lower.7ex\hbox{E}\kern-.125emX}}
\begin{document}
\title{A Generalizable 3D Diffusion Framework for Low-Dose and Few-View Cardiac SPECT}
\author{Huidong Xie, Weijie Gan, Wei Ji, Xiongchao Chen, Alaa Alashi,\\ Stephanie L. Thorn, Bo Zhou, Qiong Liu, Menghua Xia, Xueqi Guo, \\ Yi-Hwa Liu, Hongyu An, Ulugbek S. Kamilov, Ge Wang, Albert J. Sinusas, Chi Liu
\thanks{Corresponding author: Chi Liu.}
\thanks{Emails: \{Huidong.Xie; Chi.Liu\}@yale.edu}
\thanks{$^1$Department of Biomedical Engineering, Yale University, USA.}
\thanks{$^2$Department of Computer Science, Washington University in St. Louis, USA.}
\thanks{$^3$Department of Radiology and Biomedical Imaging, Yale University, USA.}
\thanks{$^5$Department of Radiology, Washington University in St. Louis, USA.}
\thanks{$^6$Department of Electrical \& Systems Engineering, Washington University in St. Louis, USA.}
\thanks{$^{9}$Department of Biomedical Engineering, Rensselaer Polytechnic Institute, USA.}}
\maketitle

\begin{abstract}
Myocardial perfusion imaging using SPECT is widely utilized to diagnose coronary artery diseases, but image quality can be negatively affected in low-dose and few-view acquisition settings. Although various deep learning methods have been introduced to improve image quality from low-dose or few-view SPECT data, previous approaches often fail to generalize across different acquisition settings, limiting their applicability in reality. This work introduced DiffSPECT-3D, a diffusion framework for 3D cardiac SPECT imaging that effectively adapts to different acquisition settings without requiring further network re-training or fine-tuning. Using both image and projection data, a consistency strategy is proposed to ensure that diffusion sampling at each step aligns with the low-dose/few-view projection measurements, the image data, and the scanner geometry, thus enabling generalization to different low-dose/few-view settings. Incorporating anatomical spatial information from CT and total variation constraint, we proposed a 2.5D conditional strategy to allow the DiffSPECT-3D to observe 3D contextual information from the entire image volume, addressing the 3D memory/computational issues in diffusion model. We extensively evaluated the proposed method on 1,325 clinical $^\text{99m}$Tc tetrofosmin stress/rest studies from 795 patients. Each study was reconstructed into 5 different low-count levels and 5 different projection few-view levels for model evaluations, ranging from 1\% to 50\% and from 1 view to 9 view, respectively. Validated against cardiac catheterization results and diagnostic comments from nuclear cardiologists, the presented results show the potential to achieve low-dose and few-view SPECT imaging without compromising clinical performance. Additionally, DiffSPECT-3D could be directly applied to full-dose SPECT images to further improve image quality, especially in a low-dose stress-first cardiac SPECT imaging protocol.

%DDPET-3D produced consistent 3D reconstructions with the proposed multi-slice condition and by fixing the starting point in the reverse sampling. With the proposed denoised prior, DDPET-3D maintains overall tracer distributions with more accurate image details. DDPET-3D achieves dose-aware denoising by embedded injected dose within the network.
\end{abstract}

\begin{IEEEkeywords}
Enter about five key words or phrases in alphabetical order, separated by commas.
\end{IEEEkeywords}

\section{Introduction}
\label{sec:introduction}
Cardiovascular diseases (CVDs) are the leading cause of mortality worldwide \cite{tsao_heart_2023}. Cardiac single-photon emission computed tomography (SPECT) plays a crucial role in diagnosing CVDs \cite{dobrucki_pet_2010}, and myocardial perfusion imaging (MPI) using SPECT remains the most widely performed procedure in nuclear cardiology \cite{abbott_contemporary_2018}. Given the growing concern of radiation exposure and potentially increased cancer risks associated with SPECT scans, reducing the SPECT injection dose is desirable \cite{lin_radiation_2010}. In addition, reducing image motion artifacts and patient discomfort through shorter scans as well as enhancing image quality on cost-effective scanners would also be beneficial in clinical settings. However, reducing the injection dose and scan duration can negatively affect image quality and diagnostic performance. Therefore, reconstructing clinically acceptable images from under-sampled measurement data is an important topic. In the context of this paper, under-sampled data refers to both low-dose and few-view data.

Conventional iterative reconstruction methods fail to produce satisfactory images with under-sampled measurement data. Deep learning techniques have been actively researched to improve the quality of medical images \cite{wang_deep_2020}. Many different deep learning methods were proposed for SPECT image reconstructions/restorations under different acquisition conditions, including low-dose/low-count SPECT imaging \cite{ramon_improving_2020, aghakhan_olia_deep_2022, chen_dudocfnet_2024, du_deep_2024}, and few-view/limited-angle SPECT imaging \cite{xie_increasing_2022, xie_deep-learning-based_2023, xie_transformer-based_2023, xie_increasing_2024}. In particular, with the advancement of cutting-edge generative models \cite{Croitoru.etal2023, Kazerouni.etal2023a}, diffusion models have demonstrated strong potential for improved nuclear imaging. While diffusion model has not been widely explored in the literature for SPECT imaging applications, it has achieved promising results for PET imaging. For example, building upon the Denoising Diffusion Probabilistic Model (DDPM) \cite{ho2020denoising}, Gong \textit{et al.,} proposed to perform 2D PET image denoising with MRI as prior information \cite{gong_pet_2023}. A 3D diffusion model is desirable to process the entire image volume, but it is impracticable due to hardware memory limit. To address the 3D imaging problem with the diffusion model, our previous work \cite{xie_dose-aware_2024} introduces DDPET-3D, a dose-aware diffusion model for 3D PET image denoising at different noise levels. Diffusion models have also been investigated for other imaging modalities \cite{gao_corediff_2023,gungor_adaptive_2023, chung_score-based_2022}. However, most deep learning models were designed for a specific acquisition setting (\textit{e.g.}, low-dose or few-view), and struggle to generalize to other settings. For example, a low-dose imaging model typically fails to produce optimal results for few-view imaging, limiting their applicability in real clinical settings. The goal of this work is to introduce a diffusion framework for generalizable 3D SPECT imaging under different acquisition conditions (i.e., low-dose and few-view settings) without further network fine-tuning or re-training. Several challenges need to be addressed to achieve this goal. 

\textbf{First}, a model to consider 3D spatial information is desirable, as SPECT is intrinsically a 3D imaging modality. However, as mentioned previously, directly training a 3D diffusion model is challenging due to memory limitation. One may train a 2D diffusion and stack all the 2D slices together to generate the 3D image volumes. However, as demonstrated in our previous work \cite{xie_dose-aware_2024} and later in this paper, this approach results in severe inconsistencies between different 2D slices along the Z-axis. Several previous works have attempted to address 3D imaging challenges in diffusion models. Chung \textit{et al.}\cite{chung_solving_2023} proposed to incorporate a total variation (TV) constrain term along the z-axis to mitigate inconsistencies within each reverse sampling step in the diffusion model. Lee \textit{et al.}\cite{lee_improving_2023} proposed to employ 2 pre-trained perpendicular 2D diffusion models to remove inconsistencies between slices. However, these two methods cannot observe 3D spatial information during the training process, leading to sub-optimal results. In contrast, our previous work proposed a 2.5D conditional strategy \cite{xie_dose-aware_2024}, which allows the network to observe neighboring slices, and achieves better performance compared to the previous two methods. However, the previously proposed 2.5D conditional strategy can only observe adjacent $n$ slices for image reconstructions, leading to a restricted receptive field, particularly at the edges of the image volumes.

\textbf{Second}, another challenge in achieving generalizable SPECT image reconstructions under different acquisition conditions is the high variation of image artifacts. The images reconstructed from noisy projection data exhibit different noise patterns compared to those reconstructed from truncated projections. The noise patterns also vary across different low-dose and few-view levels. While training separate networks for different acquisition conditions is feasible, this approach is resource intensive and may not be practical due to the high variability of clinical protocols. Moreover, a model optimized for a specific setting may struggle to generalize effectively to unseen acquisition conditions, such as a low-count level that is not included in the training data. Therefore, a method to adaptively reconstruct images under different acquisition conditions is desirable. However, most of the previously proposed deep learning methods have limited generalizability to different acquisition conditions. To improve network generalizability, we previously proposed to combine multiple U-net-based \cite{ronneberger_u-net_2015} sub-networks with varying denoising power to generate optimal denoised results for any input noise levels \cite{xie_unified_2023}. In our previous low-dose PET diffusion paper \cite{xie_dose-aware_2024}, we proposed to embed the injected dose within the diffusion model to achieve dose-aware PET denoising. We also attempted to adapt the idea of dynamic convolution \cite{xie_segmentation-free_2023, li2022odconv} for dynamic PET image denoising, addressing the variation of noise levels between early and later frames caused by tracer decays \cite{xie_noise-aware_2025}. While these approaches improved adaptability to varying image noise levels, they were designed specifically for denoising tasks and did not consider the challenges under a different acquisition setting, such as few-view settings. 

In this work, we developed a diffusion framework, DiffSPECT-3D, for generalizable 3D cardiac SPECT image reconstructions under various acquisition conditions. The main methodological contributions of the proposed DiffSPECT-3D framework are as follows: \textbf{First:} incorporating 3D spatial anatomical information from CT, we proposed a revised 2.5D conditional strategy that allows the network to observe the entire 3D CT volume for 3D SPECT image reconstructions with negligible increase in memory burden. Since CT scans are usually performed with SPECT scans for attenuation correction, they are readily available. Inspired by \cite{chung_solving_2023}, the TV penalty in the slice dimension is also added within the diffusion sampling to further improve the results. \textbf{Second:} to address the challenge of high variability in image artifacts, we proposed a consistency strategy during diffusion reverse sampling to align the diffusion prediction at each time step with the under-sampled SPECT image data, projection data, and the scanner system matrix.To achieve projection-based consistency, DiffSPECT-3D integrates the conventional iterative reconstruction updates within the diffusion sampling. To achieve image-based consistency, DiffSPECT-3D employs the diffusion posterior sampling (DPS) \cite{chung2023diffusion} strategy in each diffusion sampling step. Both techniques ensure that the diffusion reverse sampling aligns with the measured projection/image data and the scanner geometry. \textbf{Lastly:} one advantage of the proposed DiffSPECT-3D is that the diffusion training process does not require paired SPECT inputs/labels with only labels needed, eliminating the need for tedious data preparation. Since the SPECT listmode data may not always be available, the ability to perform unconditional diffusion training simplifies the training process and potentially facilitates its clinical transition.
 
\section{Methods}
\begin{figure*}[!t]
\centerline{\includegraphics[width=\textwidth]{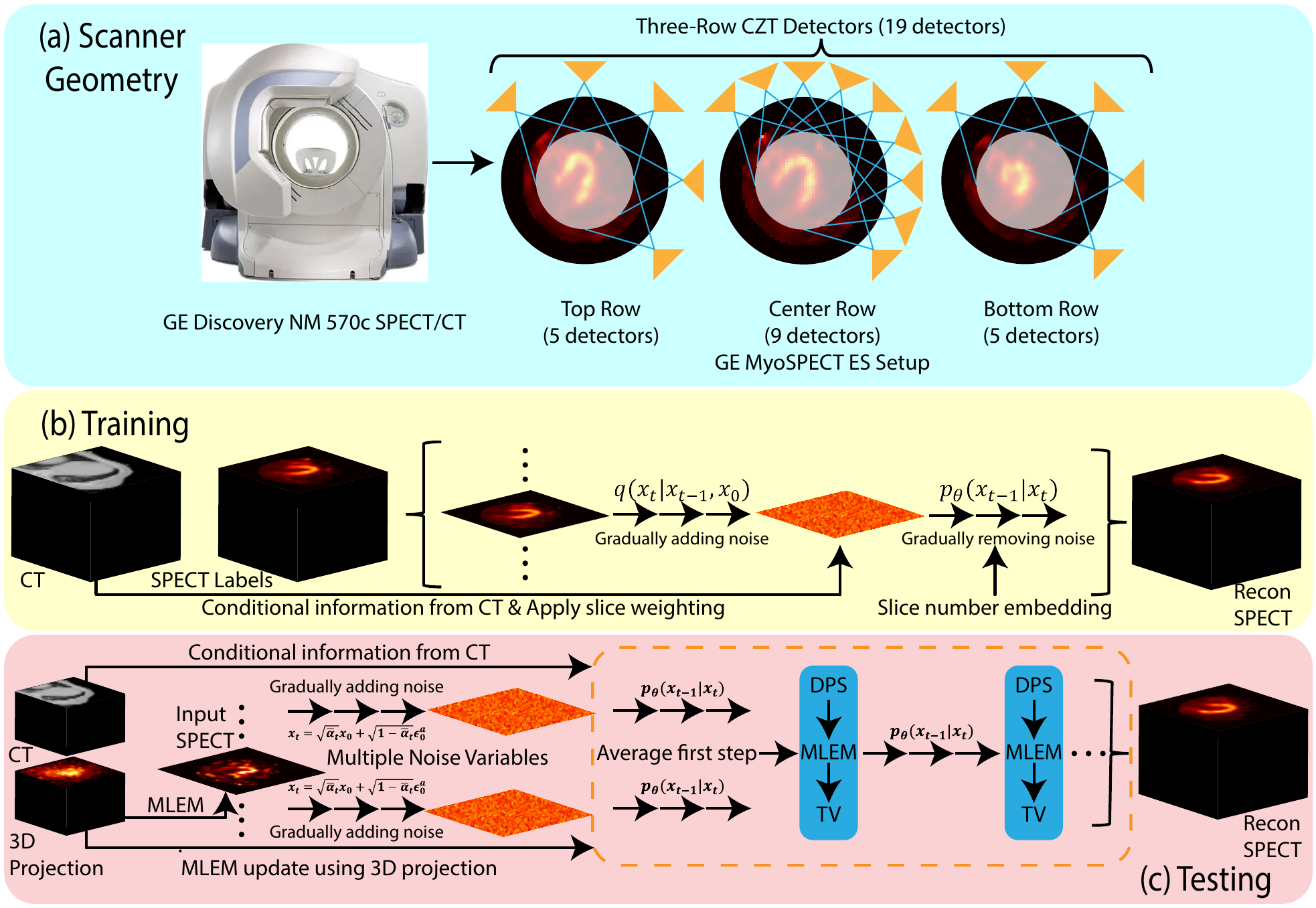}}
\caption{(a) Overview of the scanner geometry used in this work. The scanner has a total of 19 CZT detectors arranged in 3 rows. By only including the center 9 detectors, we simulated the geometry of the GE MyoSPECT ES system, a cost-effective version of the scanner. (b)-(c) Overview of the training and testing processes of the proposed DiffSPECT-3D, respectively. Training and testing procedures are detailed in Algorithms \ref{alg1}-\ref{alg2}. In summary, we proposed several techniques to enforce consistency in both the image and projection domains, enabling DiffSPECT-3D to achieve improved reconstruction quality under various acquisition conditions. Anatomical information from CT scans is also incorporated in DiffSPECT-3D for better image quality.}
\label{fig_network}
\end{figure*}

\subsection{Data Acquisition and Image Reconstruction}
A total of 1,325 clinical SPECT-CT stress/rest myocardial perfusion imaging (MPI) studies from 795 patients were included in this work. Note that not all patients have both stress and rest studies, due to the stress-first protocol. All studies were acquired at the Yale New-Haven Hospital on a GE Discovery NM/CT (DNM) 570c (GE Healthcare, Milwaukee, Wisconsin, USA) \cite{bocher_fast_2010} dedicated SPECT-CT system following the injection of $^{\text{99m}}\text{Tc-tetrofosmin}$. The collection of the patient data was approved by the institutional review board (IRB) at Yale University. All the data used in this study were de-identified prior to model training, validation, testing, and follow-up analysis.

As illustrated in Fig. \ref{fig_network}(a), arranged into 3 rows, the DNM consists of 19 cadmium zinc telluride (CZT) detector modules with tungsten pinhole collimators. Specifically, there are 5, 9, and 5 detectors placed on the top, central, and bottom rows, respectively \cite{chan_impact_2016}. All 19 detector modules are designed to simultaneously acquire 19 projections over an L-shape arc for imaging without moving the scanner. In DNM, all 19 pinhole detector modules are focused on the cardiac region to maximize imaging sensitivity over a sphere of approximately 19-cm-diameter field of view (FOV). Each detector module in DNM has full view of the heart and therefore can spend the full acquisition time to acquire photons emitted from the heart for better image quality compared to conventional dual-head SPECT systems \cite{xie_increasing_2022}. Since CZT-based systems are generally more expensive compared to conventional thallium doping NaI(Tl) cameras, it is worth mentioning the GE MyoSPECT ES system, a recent cost-effective variant of the DNM. The GE MyoSPECT ES system shares similar geometry to DNM but it only comprises the 9 central detectors instead of the three-row 19 detectors, as illustrated in Fig. \ref{fig_network}(a).

Following the clinical protocol, all images were reconstructed using the Maximum-Likelihood Expectation-Maximization (MLEM) \cite{shepp_maximum_1982, lange_em_1984} algorithm with 50 iterations. No post-filtering was applied to better visualize the improvements in image quality and the recovery of anatomical structures. The reconstruction matrix size was $50\times70\times70$ with a $4\times4\times4 \text{ mm}^3$ voxel size. The dimension of 3D projection data is $32\times32\times19$. CT scans were obtained and registered for attenuation correction and neural network training/validation/testing.

To evaluate the network performance under varying acquisition settings, images were reconstructed at low-count levels of 1\%, 5\%, 10\%, 20\%, and 50\% by equally down-sampling the listmode events. Additionally, few-view/limited-angle images were reconstructed at 1, 3, 5, 7, and 9 view by cropping the data in the projection domain. Note that the 9-view setting simulates the configurations of the GE MyoSPECT ES system. 

Although this study focuses primarily on the GE DNM and/or MyoSPECT ES systems, we believe that the proposed method could be easily extended to other imaging systems. As detailed below, this is because the network incorporates the scanner system matrix/imaging geometry within the neural network reconstruction process, and does not heavily rely on paired training inputs/labels.

\subsection{Preliminary: Diffusion Model}
The general idea of diffusion models is to learn the target data distribution $q(\xbm_0)$ (\emph{i.e.,} full-dose/full-view SPECT images in our case) using neural networks. Once the distribution is learned, we can synthesize a new sample from it. Diffusion models consist of two Markov chains: the forward diffusion process $q$ and the learned reverse diffusion process $p_\thetabm$, where $\thetabm$ denotes the learned neural network parameters.
The forward diffusion process $q$ gradually adds a small amount of Gaussian noise to $\xbm_0\sim q(\xbm_0)$ at each step, until the original image signal is completely destroyed. As defined in \cite{ho2020denoising} 
\begin{equation}
\begin{split}
q(\xbm_{1:T}|\xbm_0)&\coloneqq \prod_{t=1}^Tq(\xbm_t|\xbm_{t-1})\ , \\ \text{where} \; q(\xbm_t|\xbm_{t-1})&\coloneqq\mathcal{N}(\xbm_t;\sqrt{1-\beta_t}\xbm_{t-1},\beta_t\mathbf{I})\ .
\label{eq1}
\end{split}
\end{equation}
One property of the diffusion process is that, one can sample $\xbm_t$ for any arbitrary time-step $t$ without gradually adding noise to $\xbm_0$. By denoting $\alpha_t \coloneqq 1-\beta_t$ and $\bar{\alpha}_t \coloneqq \prod_{s=1}^t \alpha_s$, we have
\begin{equation}
\begin{split}
q(\xbm_t|\xbm_0)&=\mathcal{N}(\xbm_t;\sqrt{(\bar{\alpha}_t})\xbm_0,(1-\bar{\alpha}_t)\mathbf{I})\ , \\ 
\text{and}\ \xbm_t & = \sqrt{\bar{\alpha}_t}\xbm_0+\sqrt{1-\bar{\alpha}_t}{\bm \epsilon}\ ,
\label{eq2}
\end{split}
\end{equation}
where ${\bm \epsilon} \sim \mathcal{N}(0,\mathbf{I})$. The latent $\xbm_T$ is nearly an isotropic Gaussian distribution for a properly designed $\beta_t$ schedule. Therefore, one can easily generate a new $\xbm_T$ and then synthesize a $\xbm_0$ by progressively sampling from the reverse posterior $q(\xbm_{t-1}|\xbm_t)$.
However, this reverse posterior is tractable only if $\xbm_0$ is known
\begin{equation}
\label{equ:q-posi}
    q(\xbm_{t-1}|\xbm_t, \xbm_0) = \Ncal\Big(\xbm_{t-1}; \mu_q(\xbm_t,\xbm_0), \frac{\beta_t(1-\bar{\alpha}_{t-1})}{1-\bar{\alpha}_t}\textbf{I}\Big)\ ,
\end{equation}
where
\begin{equation}
    \mu_q(\xbm_t,\xbm_0) = \frac{\sqrt{\alpha_t}(1-\bar{\alpha}_{t-1})\xbm_t + \sqrt{\bar{\alpha}_{t-1}}(1-\alpha_t)\xbm_0}{1-\bar{\alpha}_t}\ .
    \label{eq4}
\end{equation}
Note that $q(\xbm_{t-1}|\xbm_t):=q(\xbm_{t-1}|\xbm_t, \xbm_0)$, where the extra conditioning term $\xbm_0$ is superfluous due to the Markov property.
DDPM thus proposes to learn a parameterized Gaussian transitions $p_\thetabm(\xbm_{t-1}|\xbm_t)$ to approximate the reverse diffusion posterior \eqref{equ:q-posi}
\begin{equation}
\label{equ:p-posi}
    p_\thetabm(\xbm_{t-1}|\xbm_t) = \Ncal\Big(\xbm_{t-1}; \mu_\thetabm(\xbm_t,t), \sigma^2_t\textbf{I}\Big)\ ,
\end{equation}
we can rearrange Eq \ref{eq2} to show that $x_0=\frac{x_t-\sqrt{1-\bar{\alpha}_{t}}\epsilon}{\bar{\alpha}_{t}}$, then plugging it into Eq \ref{eq4}, we can derive  $\mu_\thetabm(\xbm_t,t)$
\begin{equation}
    \label{equ:p-mean}
    \mu_\thetabm(\xbm_t,t) = \frac{1}{\sqrt{\alpha_t}}\Big(\xbm_t-\frac{1-\alpha_t}{\sqrt{1-\bar{\alpha}_t}}\epsilon_\thetabm(\xbm_t, t)\Big)\ .
\end{equation}
Here, $\epsilon_\thetabm$ denotes a neural network. 
% The training of $\epsilon_\thetabm$ is performed by optimizing the variational lower bound of the negative log likelihood. 
Through some derivations detailed in \cite{ho2020denoising}, the training objective of $\epsilon_\thetabm(\xbm_t, t)$ can be formulated as 
\begin{equation}
    \label{equ:training}
    \E_{\xbm,{\bm\epsilon},t\sim[1,T]}\big[\norm{{\bm \epsilon} - {\epsilon}_\thetabm(\xbm_t, t)}^2\big]\ .
\end{equation}
It is worth noting that the original DDPM \cite{ho2020denoising} set $\sigma_t$ to a fixed constant value based on the $\beta_t$ schedule. Recent studies~\cite{dhariwal2021diffusion,nichol2021improved} have shown improved performance by using learned variance $\sigma_t^2\coloneqq\sigma_\thetabm^2(\xbm_t, t)$. We also adopted this approach. To be specific, we have $\sigma_\thetabm(\xbm_t, t)\coloneqq\exp(v\log \beta_t + (1-v)\log\tilde{\beta}_t)$, where $\tilde{\beta}_t$ refers to the lower bound for the reverse diffusion posterior variances \cite{ho2020denoising}, and $v$ denotes the network output. We used a single neural network with two separate output channels to estimate the mean and variance of \eqref{equ:p-mean} jointly.
Based on the learned reverse posterior $p_\thetabm(\xbm_{t-1}|\xbm_t)$, the iteration of obtaining a $\xbm_0$ from a $\xbm_T$ can be formulated as follow
\begin{equation}
    \xbm_{t-1} = \mu_\thetabm(\xbm_t,t) + \sigma_t\zbm,\text{ where } \zbm\sim\Ncal(0,\textbf{I})\ .
    \label{eq8}
\end{equation}
The framework described above only allows unconditional image generations. For the purpose of low-dose/low-count SPECT image denoising or few-view/limited-angle image artifacts removal, previous works \cite{xie_dose-aware_2024, gong_pet_2023} add additional condition to the neural network. Specifically, $\epsilon_\thetabm{(\xbm_t,t)}$ becomes $\epsilon_\thetabm{(\xbm_t,t,\xbm_\mathrm{in})}$, where $\xbm_\mathrm{in}$ represents low-quality SPECT input images. However, even though this approach achieved promising results in our previous work for low-dose/low-count PET image denoising \cite{xie_dose-aware_2024}, the experimental results below show that the performance degraded significantly when applied to other acquisition conditions (e.g., few-view settings). Adding low-quality input images during network training would also require tedious data preparation steps, especially when the network targets on multiple acquisition conditions. This is even more challenging when the listmode data is not available. As detailed below, the proposed DiffSPECT-3D can be trained as an unconditional diffusion model and does not rely on paired under-sampled/fully-sampled SPECT images.

\subsection{Proposed DiffSPECT-3D}
The proposed framework is depicted in Fig. \ref{fig_network} (b)-(c). Here, we present the DiffSPECT-3D model by detailing each proposed component individually.

\subsubsection{2.5D conditional strategy}
Since SPECT is inherently a 3D imaging modality, a model to reconstruct 3D images is desirable. However, directly training a 3D diffusion model would be challenging due to the hardware memory limit. To allow the network to observe 3D information from the entire 3D image volumes during training and testing processes, DiffSPECT-3D is conditioned on the entire 3D CT image volume and aims to predict all the 2D SPECT slices sequentially. CT scans were performed for attenuation correction and are readily available in clinical settings. Specifically, the diffusion model $\epsilon_\thetabm{(\xbm_t,t)}$ becomes $\epsilon_\thetabm{(\xbm_t,t,\xbm_\mathrm{ct})}$, where $\xbm_\mathrm{ct}$ represents the CT images. Using 3D convolutional layers would significantly increase memory burden and make the network difficult to optimize. To alleviate the memory burden and allow faster convergence, we embed the entire image volumes in the channel dimension. Specifically, for a regular 2D diffusion model, the input dimension is $N_b\times W\times W\times 2$, where $N_b$ is the input batch size, $W=70$ is the width of the images, and the last dimension is the channel dimension. One channel is the conditional 2D slice, and the other one is $\xbm_{t-1}$. Using the entire 3D image volume as conditional information, the input becomes $N_b\times W\times W\times 51$ as the conditional CT images contain 50 slices. Unlike previous works \cite{xie_dose-aware_2024, gong_pet_2023}, DiffSPECT-3D does not use the under-sampled SPECT volumes as conditional information. This design prevents the network from learning information specific to a single acquisition setting, thereby enhancing its generalizability across diverse acquisition conditions. Additionally, by eliminating the need for paired SPECT images during training, DiffSPECT-3D avoids the tedious data preparation process.

To generate different 2D slices, the slice number (i.e., 1-50) is embedded within the DiffSPECT-3D network. Both sine and cosine functions were used to encode the slice number and then added with the diffusion time steps (i.e., $t+sin(\#slice)+cos(\#slice)$, where $t$ is the diffusion time-step). Note that the time-step $t$ was also encoded using both sine and cosine functions. The encoded values were then fed into 2 linear layers with Sigmoid linear unit (SiLU) in between to generate the embedding. Additionally, spatial correlation decreases as the slices are farther apart. In DiffSPECT-3D, different weight matrices $w\in\mathbb{R}^{1\times50}$ were applied along the image depth dimension when predicting different 2D slices. Specifically, when predicting the $i^{th}$ slice, other slices that are further apart from the $i^{th}$ slice are linearly scaled down based on the distance to the $i^{th}$ slice. The weight matrix can be formulated as $w(i,j)=1-\frac{|i-j|}{50}$, where $i$ and $j$ represent the predicted slice number and the slice index in the weight matrix, respectively.

\textbf{Inserting Total Variation constrain: }Even though the network can observe 3D spatial information using the proposed 2.5D conditional strategy detailed above, it still produces the image volume slice-by-slice. The experimental results showed that it caused inconsistencies along the z-axis in certain cases. To address this issue and further improve the image quality, we incorporated the TV constraint along the z-axis. Specifically, we can consider the following regularized reconstruction problem:

\begin{equation}
\hat{\xbm}=\underset{\xbm}{\text{argmin}}\frac{1}{2}||y-S\xbm||^2_2+TV(\xbm)
\label{eq12}
\end{equation}

where $y$, $\xbm$, and $S$ are the measured projection data, the reconstructed image, and the system matrix, respectively. $TV(\xbm)$ is the TV regularization term. It aims to minimize $||\textbf{D}_z(\xbm)||_1$, where $\textbf{D}_z(\xbm)$ computes the finite difference along the z-axis. Minimization of Eq \ref{eq12} can be performed using optimization algorithms, such as the alternating direction method of multipliers (ADMM) \cite{boyd_distributed_2011}. Here, unlike conventional TV algorithms, which consider the differences along all three directions, we only consider the finite difference along the z-axis. This is because the inconsistencies in the x-y plane has been already addressed by the neural network. The experimental results showed that this straightforward approach effectively addressed inconsistencies between different 2D slices and improved overall image quality.

\subsubsection{Diffusion sampling consistency using image data}
As outlined in Fig. \ref{fig_network}(c), to reconstruct the corresponding high-quality SPECT images from under-sampled inputs and to enforce an image-domain consistency, under-sampled SPECT images are incorporated during the testing phase. To achieve this, we adapted the idea of Diffusion Posterior Sampling (DPS) strategy \cite{chung2023diffusion} in the proposed DiffSPECT-3D model. By assuming the measurement noise follows a Gaussian distribution, Eq \ref{eq8} becomes:
\begin{equation}
    \xbm_{t-1} = \mu_\thetabm(\xbm_t,t) + \sigma_t\zbm - \lambda_{dps} \nabla_{\xbm_t}||\xbm_{in}-\hat{\xbm}_0||_2^2
    \label{eq11}
\end{equation}

where $\nabla_{\xbm_t}||\xbm_{in}-\hat{\xbm}_0||_2^2$ is an image data fidelity term. $\lambda_{dps}$ is a hyper-parameter controlling the weight between it and the diffusion prior. It is expected that extreme $\lambda_{dps}$ values would be problematic. The higher the $\lambda_{dps}$ values, the final network output would be more closely resemble to the input SPECT image $x_{in}$, amplifying the noise. Setting lower $\lambda_{dps}$, the network would tend to produce images with distorted anatomical structures, making the images unusable in clinical settings. In this work, $\lambda_{dps}$ is designed to be proportional to the input SPECT image count-levels. Larger $\lambda_{dps}$ values are used for higher input SPECT image count levels, allowing the diffusion sampling to rely more on the input images as higher-count inputs contain more useful information. Conversely, when the input count level is extremely low (e.g., in 1\% low-count or 1-view settings), smaller $\lambda_{dps}$ values are applied as the input images lack sufficient information and the diffusion sampling should rely more on the trained neural network. The optimal $\lambda_{dps}$ values were determined on the validation dataset for different acquisition settings. $\lambda_{dps} = 0.0698 * \ln(C) + 0.3454
$ was then fitted using MATLAB.

It is worth noting that optimal $\lambda_{dps}$ values increase as input SPECT count-levels increase, which aligns with our expectation. When the input SPECT count levels are extremely low (e.g., 1\% low-count or 1-view settings), the MLEM-reconstructed images are very noisy and contain little useful image information. As a result, high $\lambda_{dps}$ values can negatively impact the outcomes in these extremely low-count scenarios.

In the next sub-section, we propose to incorporate the iterative reconstruction updates using the projection data within the diffusion sampling steps to further improve the results, especially in extremely low-count settings, when the under-sampled SPECT images $\xbm_{in}$ contain little useful information.

\subsubsection{Diffusion sampling consistency using projection data} To further improve the results, the MLEM iterative update is inserted in each diffusion step to ensure that the diffusion prediction at each step aligns with the measured projection data and scanner geometry. Given the acquired projection data $y\in\mathbb{R}^{32\times32\times19}$, the MLEM algorithm for image reconstruction is formulated as follows: 

\begin{equation}
I_{t+1}(j)=\frac{1}{\sum_{k=1}^KS(j,k)}I_{t}(j)\sum_{k=1}^K\frac{y(k)S(j,k)}{\sum_{j=1}^JI_{t}S(j,k)} 
\end{equation}
where $j=1,2,...,J$ and $J=240,000$ $(50\times70\times70)$ denote the total number of voxels in the reconstructed image volume. $K=19,456$ $(32\times32\times19)$ is the total number of detector measurements. $I_{t}\in\mathbb{R}^{50\times70\times70}$ represents the reconstructed image volume at $t^{th}$ iteration using MLEM. The system matrix $S(j, k)\in\mathbb{R}^{245,000\times19,456}$ represents the probability for a photon emitted from the area covered by voxel $j$ to be captured by detector at location $k$.

As described above, the general idea of the diffusion model is to iteratively approximate the target data distribution $q(\xbm_0)$. In this work, we proposed to incorporate MLEM updates as a constraint within diffusion updates. MLEM provides a reliable estimate for the underlying image by leveraging the system matrix and the observed projection data, while diffusion reverse sampling serves as a powerful denoising or artifact removal algorithm to remove complex noise patterns that cannot be addressed using conventional methods such as TV regularization \cite{persson_total_2001}. Specifically, incorporating MLEM updates within diffusion sampling, the estimate of the target image $\hat{\xbm}_0$ becomes

\begin{align}
\begin{split}
    \hat{\xbm}_{0}^{mlem} = (1-\lambda_{mlem})\hat{\xbm}_0 + \lambda_{mlem}f_{mlem}(\hat{\xbm}_{0},y,S)
    \label{eq10}
\end{split}
\end{align}

where $f_{mlem}$ denotes the MLEM update. $f_{mlem}$ takes the current estimate $\hat{\xbm}_0=\frac{x_t-\sqrt{1-\bar{\alpha}_{t}}\epsilon_\theta(\xbm_t,t)}{\bar{\alpha}_{t}}$, the projection data $y$, and the system matrix $S$ as input to compute the MLEM update. $\lambda_{mlem}$ is a hyper-parameter controlling the balance between MLEM updates and the diffusion regularization term. The experimental results showed that the optimal $\lambda_{mlem}$ values are dependent on the count-levels of the input SPECT data. In this work, the optimal $\lambda_{mlem}$ values were determined based on the validation dataset for all the different acquisition settings. $\lambda_{mlem}(C)=0.1559 * \exp(-4.8120C) + 0.0079 * \exp(3.6508C)
$ was then fitted using MATLAB, where $C$ represents the SPECT data count levels. Note that even though projection data is used as a constraint in Eq. \ref{eq10}, similar to Eq. \ref{eq11}, this projection-based consistency term also operates in the image space.

It is worth noting that by implementing the DPS and/or the MLEM insertion strategies, it is feasible to employ a single unconditional diffusion model for image reconstructions under various acquisition conditions, without additional network fine-tuning or re-training. Furthermore, the network can be trained solely on fully-sampled SPECT images and does not require paired low-quality/high-quality images for training, thus bypassing the tedious data preparation steps. In this work, MLEM update is performed every 10 diffusion steps to reduce the overall sampling time.

Based on the fitted $\lambda_{dps}$ and $\lambda_{mlem}$, we observed that as the input SPECT data count levels decrease, the network increasingly relies on the projection-based consistency term (i.e., MLEM insertion) over the image-based consistency term (i.e., DPS strategy). This is the same as our expectation. In extremely low-count settings, the highly noisy SPECT images contribute less effectively to the final network outputs, and the network tends to rely more on the original projection data, which is less nosier compared to the iteratively-reconstructed images. Conversely, with high-count inputs, the network can extract more useful information from the reconstructed images, thus giving higher weight to the image-domain consistency term.

\subsubsection{Diffusion sampling starting point}
Predicting high-quality SPECT images from under-sampled measurement data is an ill-posed problem, and we could generate an infinite number of different reconstructed images using different starting Gaussian point (i.e., $\xbm_T \sim \Ncal(0,\textbf{I})$). To address this issue and improve image quality, in DiffSPECT-3D, the diffusion reverse sampling starts from the under-sampled SPECT images $\xbm_{in}$. Specifically, in DiffSPECT-3D, the reverse sampling starts from $\xbm_T  = \sqrt{\bar{\alpha}_T}\xbm_{in}+\sqrt{1-\bar{\alpha}_T}{\bm \epsilon}$. The intuition is to fix the forward and reverse processes in diffusion sampling. Experimental results showed that it helps producing images with reduced imaging artifacts and improved quantitative measurements.

\textbf{In summary}, DiffSPECT-3D has the following contributions: (1) By incorporating CT scans, the DiffSPECT-3D can observe 3D spatial anatomical information with the proposed 2.5D conditional strategy, without introducing excessive memory burden. To reduce inconsistencies between different reconstructed 2D slices, a total variation regularization term is added within the reverse sampling to address the inconsistency issue between different 2D slices and to further reduce image noise. (2) We adapted the DPS strategy to enforce the consistency between the diffusion prediction at each time step and the under-sampled image data. (3) Additionally, we proposed to incorporate MLEM updates within diffusion sampling steps to align diffusion predictions with the measured projection data and the scanner system matrix. With (2) and (3), DiffSPECT-3D does not require paired SPECT inputs/labels for training and achieves image reconstruction under various acquisition conditions, without further network fine-tuning or retraining. (4) Instead of complete Gaussian noise, the under-sampled SPECT images are used as the starting point of diffusion sampling, further improving the image quality.

It should be noted that a similar 2.5D conditional strategy was introduced in our previous work \cite{xie_dose-aware_2024}. The difference is that in that work the network was only conditioned on neighboring $n$ slices. But in DiffSPECT-3D, the entire image volume is used as conditional information, and the slice number embedding and a weighting strategy are introduced to reconstruct different 2D slices. This approach allows the network to have a larger receptive field, which enhances its ability to capture more contextual information.

In addition to these, other techniques proposed and demonstrated to be useful in our previous work \cite{xie_dose-aware_2024} are also incorporated in DiffSPECT-3D. Specifically, during the testing stage, the added Gaussian noise $\epsilon$ is fixed for all 2D slices in a 3D volume. This ensures that reverse sampling converges to the same location in the target domain $q(\xbm_0)$, addressing the ill-posed nature of the reconstruction problem, while only conditional information changes to reconstruct different 2D slices. In addition, in DiffSPECT-3D, reverse sampling starts from two different noise variables $\epsilon_0^a$ and $\epsilon_0^b$. This is to address the artifacts dependent on the same Gaussian distribution, as detailed in \cite{xie_dose-aware_2024}. Algorithm \ref{alg1} displays the complete training procedure of the proposed DiffSPECT-3D. Algorithm \ref{alg2} displays the complete sampling procedure of the proposed DiffSPECT-3D.

\textbf{Implementation details: }All the network-related programs were implemented using PyTorch with Python. SPECT list-mode processing and image reconstructions were implemented using C++ and MATLAB, respectively. A GPU-based MLEM algorithm was implemented with PyTorch to accelerate diffusion sampling. DDIM sampling \cite{song_denoising_2022} with 25 steps was used in DiffSPECT-3D and all related experiments in this work, unless otherwise specified. Training and testing were conducted on an NVIDIA H100 GPU. The training took approximately 5 days to fully converge. By paralleling the predictions of all the 2D slices, DiffSPECT-3D only took roughly 2.1 seconds to generate a 3D SPECT volume.

\begin{algorithm}[t]
\caption{Training}\label{alg1}
\begin{algorithmic}
\State \textbf{Repeat}
\State $\xbm_0\sim q(\xbm_0)$ \Comment{Sample single-slice fully-sampled data}
\State $\xbm_{\mathrm{ct}}\sim q(\xbm_{\mathrm{ct}})$ \Comment{Sample 3D CT volumes}
\State $t\sim \mathrm{Uniform({1,...,T})}$ \Comment{Sample diffusion time-step}
\State ${\bm \epsilon} \sim \mathcal{N}(0,\mathbf{I})$
\State $\nabla_\theta||{\bm \epsilon}-\epsilon_\thetabm(\sqrt{\bar{\alpha}_t}\xbm_0+\sqrt{1-\bar{\alpha}_t}{\bm \epsilon},t,\xbm_\mathrm{ct})$ \\ \hfill \Comment{Take gradient descent step}
\State \textbf{Until} convergence
\end{algorithmic}
\end{algorithm}

\begin{algorithm}
\caption{Testing}\label{alg2}
\begin{algorithmic}
\State $y$ \Comment{3D under-sampled projection}
\State $S$ \Comment{Scanner system matrix}
\State $\xbm_{\mathrm{in}}\sim q(\xbm_{\mathrm{in}})$ \Comment{Get 3D under-sampled image volume}
\State $\xbm_{\mathrm{ct}}\sim q(\xbm_{\mathrm{ct}})$ \Comment{Get 3D CT image volume}

\State ${\bm \epsilon}_0\sim \mathcal{N}(0,\mathbf{I})$ \Comment{Get noise variables}

\State $\xbm_{T} = \sqrt{\bar{\alpha}_{T}}\xbm_{\mathrm{in}}+\sqrt{1-\bar{\alpha}_{T}}{\bm \epsilon}_{0}$
\\
\While{$t=T,...,0$}

\While{$i=1,...,I$}

$\xbm_{t-1}[i]=\mathrm{Sampler}(\xbm_{t}[i],\xbm_{\mathrm{in}},\xbm_{\mathrm{ct}},i)$

\Comment{sampling for all 2D slices}

\EndWhile

$\xbm_{t-1}=\text{DPS}(\xbm_{t-1})$ \Comment{DPS based on Eq \ref{eq11}}

$\xbm_{t-1}=\text{MLEM}(\xbm_{t-1}, y, S)$ \Comment{MLEM based on Eq \ref{eq10}}

$\xbm_{t-1}=\text{TV}(\xbm_{t-1})$ \Comment{TV based on Eq \ref{eq12}}

\EndWhile

\State \textbf{Return }$\xbm_{0}$
\end{algorithmic}
\end{algorithm}

\section{Results}
In the following sub-sections, representative cases were selected and presented to show that the proposed DiffSPECT-3D could be directly applied to different acquisition settings without further network fine-tuning or re-training. Comparisons with other related methods and ablation experiments were also conducted and included in this section. For each patient, the normal/abnormal status was determined based on clinical diagnostic reports and cardiac catheterization results (if available).

\subsection{Low-dose/low-count settings}

\begin{figure*}[!t]
\centerline{\includegraphics[width=\textwidth]{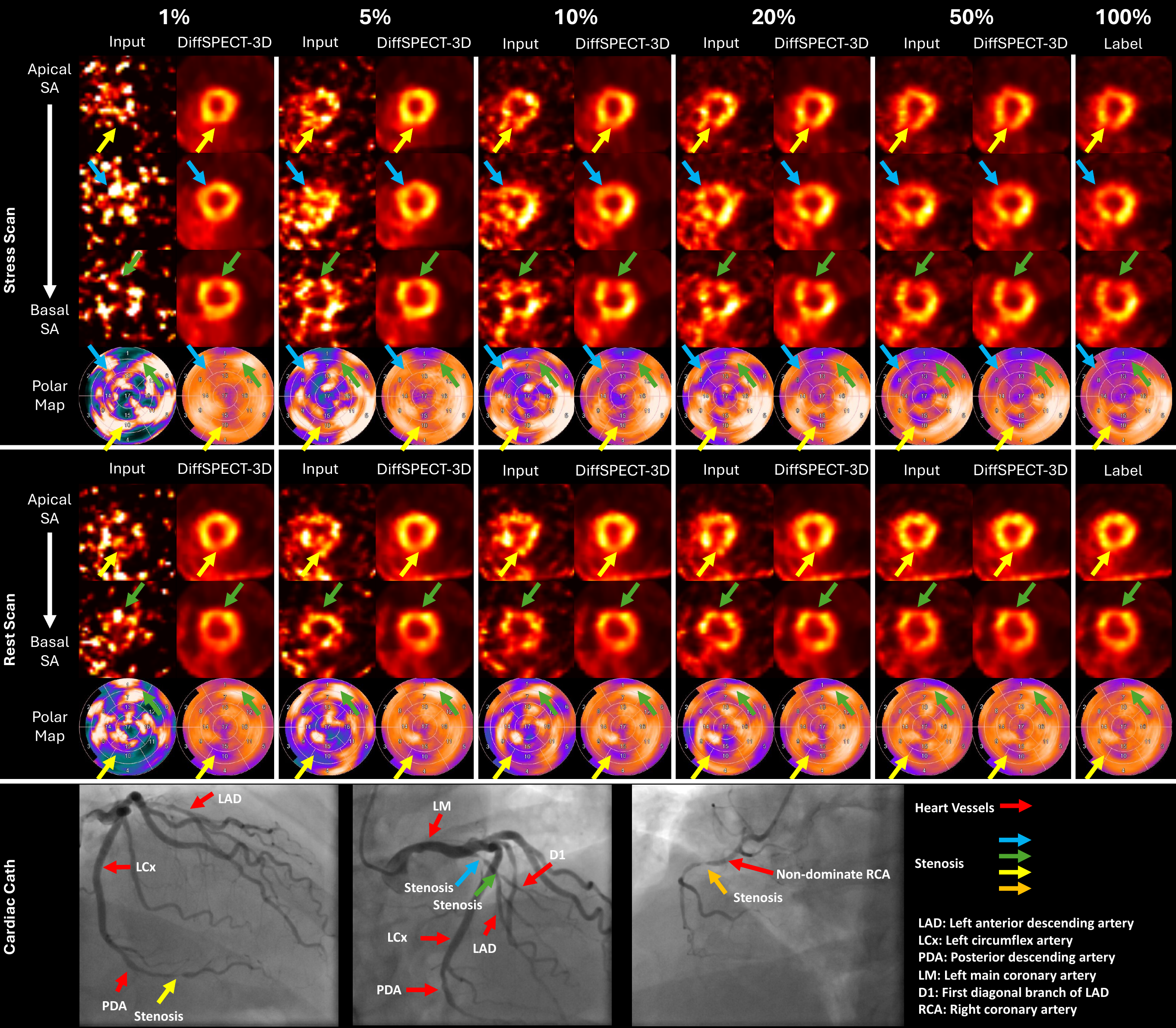}}
\caption{A representative patient study reconstructed using the proposed DiffSPECT-3D with different low-count levels. Red arrows in the cardiac catheterization images point to different coronary vessels. Non-red arrows point to different perfusion defects in this patient. Arrows with the same color correspond to the same perfusion defect. This patient has a large and critical stenosis in the proximal LAD (blue arrows), with an additional stenosis in the diagonal LAD (green arrows) and diffuse distal LAD disease. As pointed by the yellow arrows, this patient also has a stenosis in the PDA of the dominant LCx. The stress and rest images show that the large anteroseptal defect (blue arrows) was reversible with some residual scar in the anterolateral wall within that defect (green arrows). Also, based on the stress and rest results, the mid-inferior wall defect is primarily fixed (yellow arrows). SA: short-axis.}
\label{fig_low_dose_results}
\end{figure*}

A representative patient study was chosen and presented in Fig. \ref{fig_low_dose_results}. This patient is a 54-year-old male with a history of coronary artery disease. Based on clinical diagnosis and confirmed by cardiac catheterization study, this is a highly abnormal SPECT myocardial perfusion study following exercise at an average workload for the patient's age and gender. In this paper, catheterization was used to determine the presence/absence of obstructive lesions corresponding to the locations shown in SPECT images. We consider catheterization as the gold standard for the defect information for human studies. 

The perfusion imaging of this patient showed a large sized, moderate to severe intensity, reversible perfusion defects in the basal to apical anterior (green arrows in Fig. \ref{fig_low_dose_results}), and basal to apical anteroseptal (blue arrows in Fig. \ref{fig_low_dose_results}) walls consistent with ischemia. This patient also has a primarily fixed mid-inferior wall defect (yellow arrows in Fig. \ref{fig_low_dose_results}) consistent with ischemia. The proposed DiffSPECT-3D showed promising denoising performance in various low-count settings. DiffSPECT-3D accurately recovered the perfusion defects at certain low-count levels (above 20\% for both stress and rest studies for this patient) and the denoised images are consistent with the diagnostic comments and the cardiac catheterization results. At extremely low-count settings (1\% to 10\%), DiffSPECT-3D can still reconstruct the overall contour of the myocardium, even though some perfusion defects were not accurately recovered.

\subsection{Few-view/limited-angle settings}
\begin{figure*}[!t]
\centerline{\includegraphics[width=\textwidth]{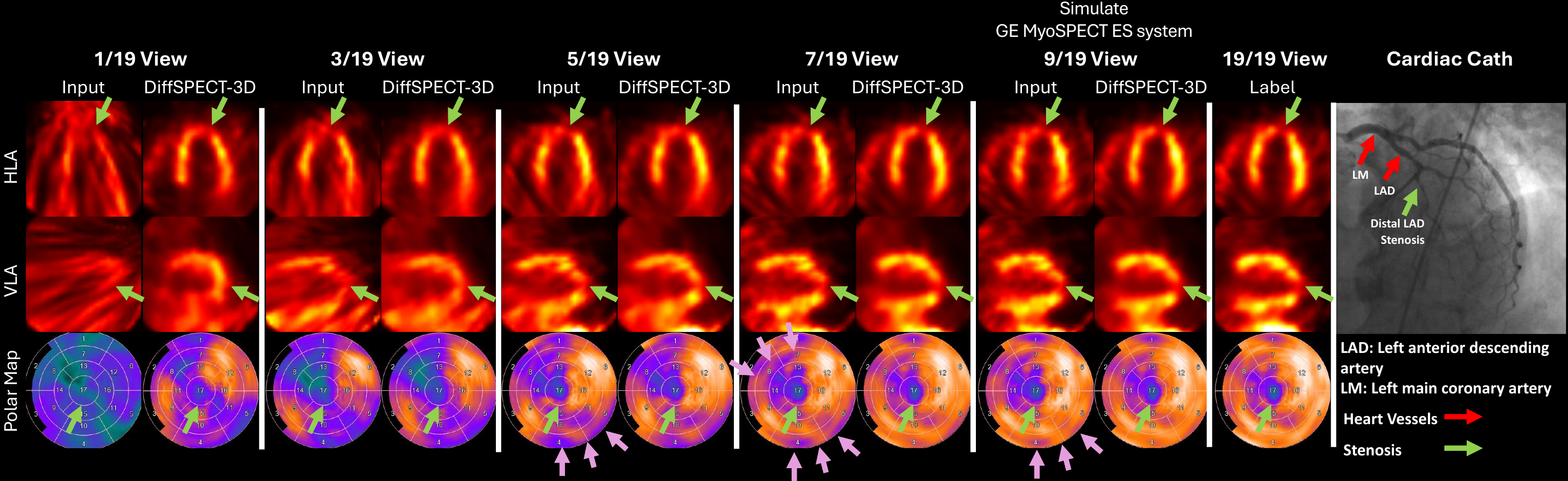}}
\caption{A representative patient study reconstructed using the proposed DiffSPECT-3D with different few-view settings. Red arrows in the cardiac catheterization images point to different coronary vessels. Non-red arrows point to the perfusion defect in this patient. Arrows with the same color correspond to the same perfusion defect. The cardiac catheterization image shows that this patient has a stenosis in the distal LAD (green arrows), leading to medium sized, mild to moderate intensity perfusion defects in the apex, apical inferior, apical anterior, and apical lateral walls. Pink arrows point to few-view artifacts that were effectively suppressed by the proposed DiffSPECT-3D. The 9-view setting simulates a GE MyoSPECT ES system. HLA: horizontal long-axis. VLA: vertical long-axis. Stress SPECT images are presented.}
\label{fig_few_view_results}
\end{figure*}

Another representative patient study was chosen and presented in Fig. \ref{fig_few_view_results}. This patient is a 77-year-old female with a history of hyperlipidemia, hypertension, aortic valve stenosis, cardiomyopathy of unknown etiology with mixed diastolic and systolic heart failure. Diagnostic comments from nuclear medicine physician showed that this is a abnormal SPECT myocardial perfusion study showing a medium-sized, mild to moderate intensity, reversible perfusion defect in the apex,  apical inferior, apical anterior, and apical lateral walls consistent with ischemia. At 7- and 9-view settings, the proposed DiffSPECT-3D effectively suppressed few-view imaging artifacts, and the true apical wall perfusion defects become better visualized in the polar maps and the SPECT images (green arrows in Fig. \ref{fig_few_view_results}). At extremely few-view settings (1-, 3-, and 5-view), DiffSPECT-3D was still able to recover the overall contour of the myocardium, even though the few-view imaging artifacts were not able to be completely removed.

The presented results in the above 2 sub-sections showed the potential of achieving low-count/low-dose/few-view/limited-angle SPECT imaging without compromising the image quality and clinical performance, though a further clinical validation is needed on a larger-scale dataset with clinical gold standard. The presented results in Fig. \ref{fig_few_view_results} also show that, with the proposed DiffSPECT-3D, the 9-view GE MyoSPECT ES system could achieve similar image quality to the more expensive 19-view GE Alcyone 570c SPECT system.

\subsection{Denoise Full-dose/Full-view Stress studies}

\begin{figure*}[!t]
\centerline{\includegraphics[width=\textwidth]{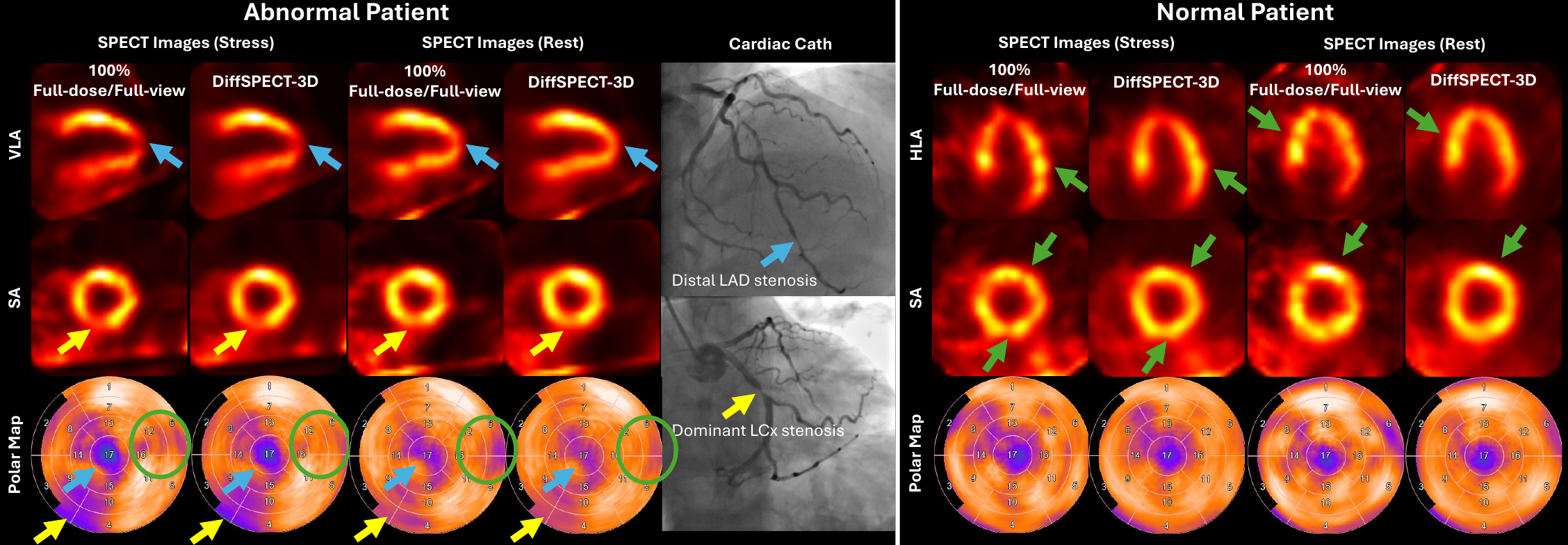}}
\caption{One abnormal and one normal patient studies reconstructed using the proposed DiffSPECT-3D using full-dose/full-view images as input. The normal/abnormal status was determined based on clinical diagnostic reports. Non-red arrows in the cardiac catheterization images point to the perfusion defect in the abnormal patient. Arrows with the same color correspond to the same perfusion defect. The cardiac catheterization image shows that the abnormal patient has a stenosis in the distal LAD (blue arrows), and another stenosis in the dominant LCX (yellow arrows). For the normal patient, the lower uptake apical wall may be attributed to apical thinning, as noted in the diagnostic report. Catheterization was not performed for the normal patient. Green arrows/circles indicate different noisy artifacts that were effectively suppressed by the proposed DiffSPECT-3D. HLA: horizontal long-axis. VLA: vertical long-axis. SA: short-axis. LAD: left anterior descending artery. LCx: left circumflex artery.}
\label{fig_stress_denoise_results}
\end{figure*}

At the Yale New-Haven Hospital, patients usually undergo a stress scan first followed by a rest scan. By performing the stress scan first, if the stress images show normal myocardial perfusion, the rest scan may not be necessary in some cases, reducing the overall radiation exposure and improving patient throughput. Furthermore, a \textbf{ low-dose stress / high-dose rest} $\text{Tc-99m}$ protocol is implemented at Yale to further reduce the overall radiation exposure. Therefore, denoising stress SPECT images would be a meaningful application in reality, improving image quality in a clinical low-dose setting.

To show the proposed DiffSPECT-3D can be adopted for this task, we tried to apply the trained model on full-dose/full-view stress SPECT images. One abnormal and one normal patient studies are presented in Fig. \ref{fig_stress_denoise_results}.

The abnormal patient in Fig. \ref{fig_stress_denoise_results} is a 63-year-old male. As confirmed by nuclear cardiologist, the perfusion imaging of this patient showed two, non-contiguous defects. One small sized, mild intensity, reversible perfusion defect in the apical wall (blue arrows in Fig. \ref{fig_stress_denoise_results}) is consistent with ischemia; and one small sized, mild intensity, reversible perfusion defect in the basal to mid inferoseptal wall (yellow arrows in Fig. \ref{fig_stress_denoise_results}) is consistent with ischemia. As shown in the cardiac catheterization images, the distal left anterior descending artery (LAD) stenosis led to the apical defect, and the dominant left circumflex (LCX) stenosis led to the basal to mid inferoseptal defect. The proposed DiffSPECT-3D effectively suppressed noisy artifacts in the images (e.g., green circles in Fig. \ref{fig_stress_denoise_results}), and the two perfusion defects are clearly visualized. We also tested the network on full-dose/full-view rest images from this patient, and the presented results demonstrated DiffSPECT-3D's generalizability to image denoising of different noise-levels.

The normal patient in Fig. \ref{fig_stress_denoise_results} is a 50-year-old female. As confirmed by nuclear cardiologist, the perfusion imaging of this patient is normal. The reduced radiotracer uptake in the apex of this patient may be attributed to apical thinning, as noted in the diagnostic report. Apical thinning can make the apex appear thinner compared to other regions of the myocardium. This is a normal anatomical finding and is often considered an image artifact rather than a true indication of heart disease \cite{dvorak_interpretation_2011}. The proposed DiffSPECT-3D effectively suppressed noisy artifacts in images (e.g., green arrows in Fig. \ref{fig_stress_denoise_results}) and produced a more uniform myocardium for this normal patient. Note that since cardiac catheterization for this patient was not available, it is not certain whether this is a true apical thinning. Nevertheless, the proposed DiffSPECT-3D produced lower-noise images without affecting this lower-uptake region.

\subsection{Ablation Studies}

\begin{table*}[!h]
\centering
\caption{Quantitative assessment for ablated methods. The measurements were obtained by averaging the values on the testing human studies. The proposed method consistently produced promising reconstruction results regardless of input count levels and few-view settings. $\uparrow$ and $\downarrow$ indicate that higher and lower values are better, respectively, for each metric.}
\resizebox{\textwidth}{!}{
\begin{tabular}{c|c|c|c|c|c}
\hline
\multicolumn{6}{c}{\textbf{Low-dose/Low-count settings}}\\
\hline\hline
     \textbf{PSNR$\uparrow$/NRMSE$\downarrow$/SSIM$\uparrow$} &  1\% Count Input&   5\% Count Input&    10\% Count Input&  20\% Count Input& 50\% Count Input\\

\hline
Input  &17.439 / 0.153 / 0.707 & 26.402 / 0.053 / 0.834 & 30.001 / 0.035 / 0.892 & 33.738 / 0.023 / 0.940 & 39.485 / 0.011 / 0.979  \\
\hline
DiffSPECT-3D (w/o $\xbm_{in}$ start, DPS, TV, MLEM) & 28.761 / 0.038 / 0.892 & 29.251 / 0.036 / 0.899 & 29.291 / 0.036 / 0.900 & 29.307 / 0.036 / 0.900 & 29.314 / 0.036 / 0.900  \\
 \hline
DiffSPECT-3D (w/o DPS, TV, MLEM) & 29.454 / 0.036 / 0.900 & 30.544 / 0.032 / 0.911 & 30.636 / 0.032 / 0.912 & 30.668 / 0.032 / 0.913 & 30.678 / 0.032 / 0.913 \\
 \hline
DiffSPECT-3D (w/o TV, MLEM)  & 29.454 / 0.036 / 0.900 & 31.247 / 0.030 / 0.915 & 32.935 / 0.024 / 0.932 & 34.925 / 0.019 / 0.951 & 39.232 / 0.012 / 0.978  \\
\hline
DiffSPECT-3D (w/o MLEM)  & 29.502 / 0.035 / 0.901 & 31.471 / 0.029 / 0.916 & 33.289 / 0.023 / 0.934 & 35.296 / 0.018 / 0.953 &  39.432 / 0.011 / 0.979   \\
\hline
\hline
\textbf{DiffSPECT-3D (proposed)}  & \textcolor{red}{29.988 / 0.033 / 0.901} & \textcolor{red}{31.878 / 0.027 / 0.920} & \textcolor{red}{33.425 / 0.023 / 0.936} & \textcolor{red}{35.378 / 0.011 / 0.979} & \textcolor{red}{39.853 / 0.011 / 0.982}  \\
\hline

\multicolumn{6}{c}{\textbf{Few-view/Limited-angle settings}}\\
\hline\hline
     \textbf{PSNR$\uparrow$/NRMSE$\downarrow$/SSIM$\uparrow$} &  1/19 View &   3/19 View & 5/19 View &  7/19 View & 9/19 View\\

\hline
Input  &25.066 / 0.060 / 0.804 & 27.893 / 0.043 / 0.868 & 30.591 / 0.031 / 0.907 & 32.304 / 0.025 / 0.930 & 33.337 / 0.023 / 0.941  \\
\hline
DiffSPECT-3D (w/o $\xbm_{in}$ start, DPS, TV, MLEM) & 29.334 / 0.036 / 0.900 & 29.330 / 0.036 / 0.900 & 29.322 / 0.036 / 0.900 & 29.314 / 0.036 / 0.900 & 29.308 / 0.036 / 0.900  \\
 \hline
DiffSPECT-3D (w/o DPS, TV, MLEM) & 29.328 / 0.036 / 0.898 & 29.896 / 0.034 / 0.907 & 30.149 / 0.033 / 0.908 & 30.170 / 0.033 / 0.909 & 30.199 / 0.033 / 0.909 \\
 \hline
DiffSPECT-3D (w/o TV, MLEM)  & 29.454 / 0.035 / 0.899 & 30.099 / 0.034 / 0.908 & 31.490 / 0.028 / 0.925 & 32.717 / 0.024 / 0.937 & 33.524 / 0.022 / 0.945  \\
\hline
DiffSPECT-3D (w/o MLEM)  & 29.379 / 0.036 / 0.901 & 30.033 / 0.033 / 0.909 & 31.609 / 0.027 / 0.926 & 32.837 / 0.024 / 0.938 &  33.643 / 0.022 / 0.945   \\
\hline
\hline
\textbf{DiffSPECT-3D (proposed)}  & \textcolor{red}{29.975 / 0.034 / 0.906} & \textcolor{red}{30.078 / 0.033 / 0.909} & \textcolor{red}{31.634 / 0.027 / 0.926} & \textcolor{red}{32.888 / 0.024 / 0.938} & \textcolor{red}{33.702 / 0.022 / 0.946}  \\
\hline
\end{tabular}
}
\label{table1}
\end{table*}

\begin{figure*}[!t]
\centerline{\includegraphics[width=0.8\textwidth]{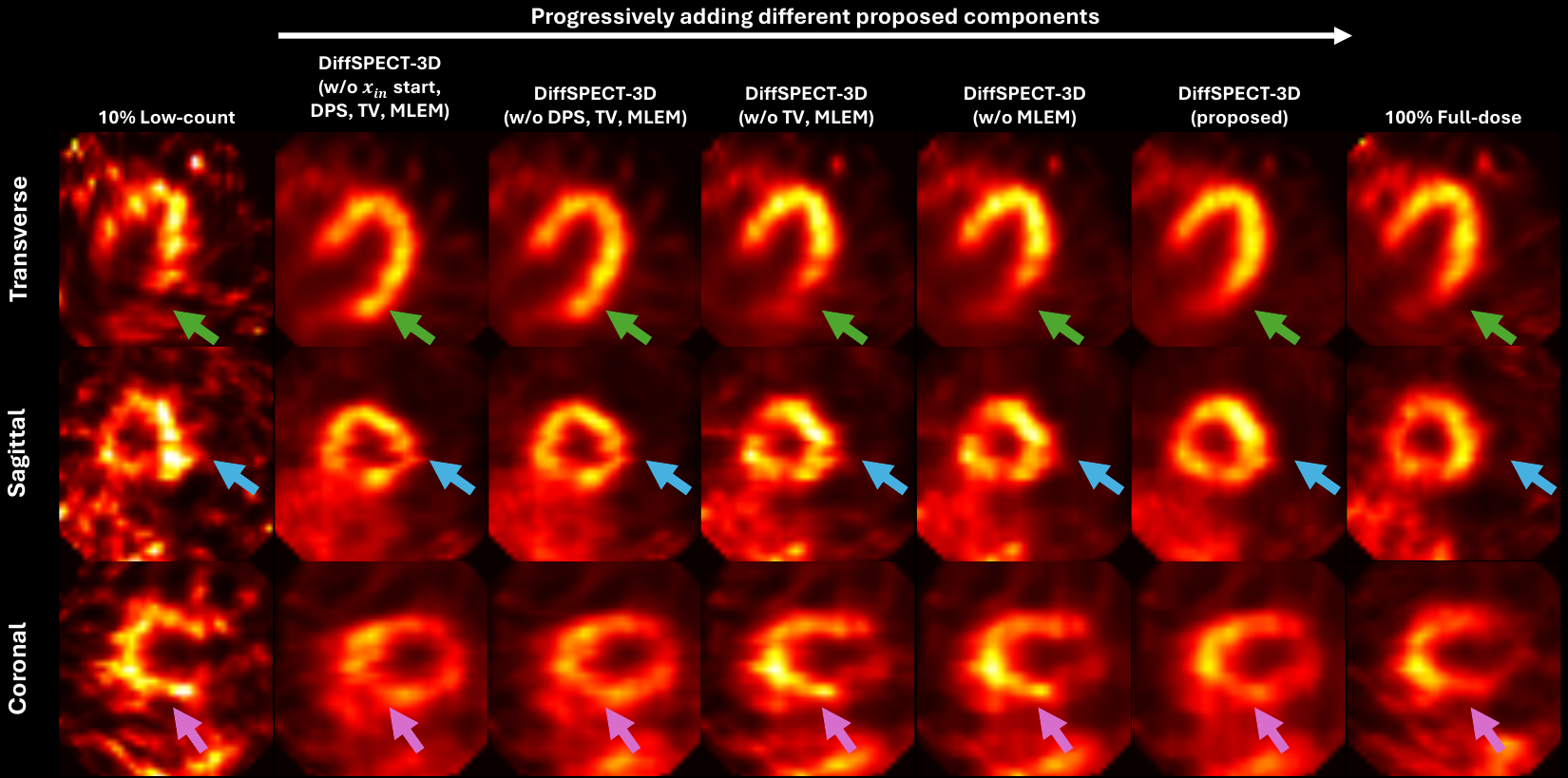}}
\caption{Results from a testing case to demonstrate the effectiveness of different proposed components in DiffSPECT-3D. Arrows point to regions that were effectively recovered using the proposed components.}
\label{fig_ablation_study}
\end{figure*}

We performed several ablated experiments to demonstrate the effectiveness of different proposed components in the DiffSPECT-3D. The image quality was quantified using PSNR (peak signal-to-noise ratio), NRMSE (Normalized Root Mean Square Error), and SSIM (Structural Similarity). All the quantitative measurements are included in Table \ref{table1}. As illustrated in Fig. \ref{fig_ablation_study}, different proposed techniques were progressively added to visualize their effectiveness.

\textbf{Impact of Denoising Prior: } By comparing the second and the third columns in Fig. \ref{fig_ablation_study}, it can be observed that the proposed denoising prior could help suppress undesirable artifacts in the images, as indicated by the blue arrows in the sagittal view. Note that without the denoising prior, the "DiffSPECT-3D (w/o prior, DPS, TV, MLEM)" network does not have low-quality SPECT images as input. \textbf{Impact of Diffusion Posterior Sampling: } Without the DPS strategy, the image reconstructions were only based on CT images, leading to distorted myocardium and anatomical structures at certain regions, highlighted by the green arrows in the transverse view. Comparing the third and the fourth columns in Fig. \ref{fig_ablation_study}, networks with DPS generated images with better recovery of myocardium contours and anatomical details, demonstrating the effectiveness of this image-based consistency strategy. \textbf{Impact of Total Variation Constrain: } Even though the network can observe 3D spatial information using the 2.5D conditional strategy, it still introduced inconsistencies in certain regions, as shown by the blue arrows in the sagittal view. Comparing the fourth and fifth columns in Fig. \ref{fig_ablation_study}, the addition of a TV constrain within each diffusion sampling steps effective suppressed undesirable noise and inconsistencies between different slices. \textbf{Impact of Inserting MLEM Update: } While image-based consistency using DPS improves the results, it may introduce noisy artifacts from the under-sampled images into the network output. Comparing the fifth and sixth columns in Fig. \ref{fig_ablation_study}, inserting MLEM updates could suppress these undesirable artifacts, as indicated by the pink arrows in the coronal view.

As shown in Table \ref{table1}, the quantitative measurements gradually improved by incorporating different proposed components in DiffSPECT-3D.

\subsection{Comparison with Other Methods}
\begin{table*}[!h]
\centering
\caption{Comparison between different deep learning methods. The measurements were obtained by averaging the values on the testing human studies. The proposed method consistently produced promising reconstruction results regardless of input count levels and few-view settings. $\uparrow$ and $\downarrow$ indicate that higher and lower values are better, respectively, for each metric.}
\resizebox{\textwidth}{!}{
\begin{tabular}{c|c|c|c|c|c}
\hline
\multicolumn{6}{c}{\textbf{Low-dose/Low-count settings}}\\
\hline\hline
     \textbf{PSNR$\uparrow$/NRMSE$\downarrow$/SSIM$\uparrow$} &  1\% Count Input&   5\% Count Input&    10\% Count Input&  20\% Count Input& 50\% Count Input\\

\hline
Input  &17.439 / 0.153 / 0.707 & 26.402 / 0.053 / 0.834 & 30.001 / 0.035 / 0.892 & 33.738 / 0.023 / 0.940 & 39.485 / 0.011 / 0.979  \\
\hline
UNet w/o CT & 29.385 / 0.038 / 0.892 & 30.213 / 0.031 / 0.902 & 30.892 / 0.028 / 0.913 & 32.283 / 0.026 / 0.920 & 32.979 / 0.023 / 0.930\\
\hline
UNet & 29.887 / 0.037 / 0.897 & 30.828 / 0.030 / 0.913 & 31.121 / 0.025 / 0.918 & 32.440 / 0.023 / 0.923 & 33.124 / 0.016 / 0.933\\
\hline
DDIM-SPECT w/o CT & 27.096 / 0.046 / 0.841 & 27.811 / 0.042 / 0.855 & 27.930 / 0.042 / 0.858 & 27.984 / 0.042 / 0.860 & 28.057 / 0.041 / 0.860 \\
\hline
DDIM-SPECT & 28.732 / 0.038 / 0.887 & 28.931 / 0.037 / 0.890 & 28.953 / 0.037 / 0.890 & 28.967 / 0.037 / 0.891 & 28.990 / 0.037 / 0.891  \\
\hline
DiffusionMBIR w/o CT & 28.869 / 0.037 / 0.883 & 29.511 / 0.035 / 0.893 & 29.617 / 0.034 / 0.895 & 29.689 / 0.034 / 0.896 & 29.751 / 0.034 / 0.897 \\
\hline
DiffusionMBIR & 29.174 / 0.036 / 0.899 & 29.842 / 0.034 / 0.908 & 29.862 / 0.034 / 0.908 & 29.880 / 0.034 / 0.908 & 29.897 / 0.034 / 0.908 \\ 
\hline
DDSPECT-3D w/o CT & 28.546 / 0.039 / 0.891 & 30.007 / 0.032 / 0.913 & 31.180 / 0.029 / 0.924 & 31.220 / 0.029 / 0.924 & 31.256 / 0.029 / 0.924 \\
\hline
DDSPECT-3D & 29.202 / 0.037 / 0.900 & 31.103 / 0.029 / 0.918 & 31.411 / 0.028 / 0.920 & 31.594 / 0.028 / 0.921 & 31.689 / 0.027 / 0.922 \\
\hline
\hline
\textbf{DiffSPECT-3D w/o CT}  & 28.572 / 0.039 / 0.871 & 30.853 / 0.030 / 0.902 & 32.740 / 0.024 / 0.924 & 34.738 / 0.019 / 0.942 & 37.625 / 0.014 / 0.962  \\
\hline
\textbf{DiffSPECT-3D (proposed)}  & \textcolor{red}{29.988 / 0.033 / 0.901} & \textcolor{red}{31.878 / 0.027 / 0.920} & \textcolor{red}{33.425 / 0.023 / 0.936} & \textcolor{red}{35.378 / 0.011 / 0.979} & \textcolor{red}{39.853 / 0.011 / 0.982}  \\
\hline

\multicolumn{6}{c}{\textbf{Few-view/Limited-angle settings}}\\
\hline\hline
     \textbf{PSNR$\uparrow$/NRMSE$\downarrow$/SSIM$\uparrow$} &  1/19 View &   3/19 View & 5/19 View &  7/19 View & 9/19 View\\

\hline
Input  &25.066 / 0.060 / 0.804 & 27.893 / 0.043 / 0.868 & 30.591 / 0.031 / 0.907 & 32.304 / 0.025 / 0.930 & 33.337 / 0.023 / 0.941  \\
\hline
UNet w/o CT & 29.356 / 0.041 / 0.891 & 29.515 / 0.038 / 0.898 & 30.496 / 0.033 / 0.906 & 30.895 / 0.029 / 0.915 & 31.329 / 0.027 / 0.935\\
\hline
UNet & 29.563 / 0.039 / 0.897 & 29.726 / 0.035 / 0.908 & 30.874 / 0.031 / 0.915 & 31.026 / 0.028 / 0.925 & 32.524 / 0.027 / 0.938\\
\hline
DDIM-SPECT w/o CT & 26.106 / 0.052 / 0.813 & 27.160 / 0.046 / 0.842 & 27.718 / 0.043 / 0.855 & 27.891 / 0.042 / 0.859 & 27.889 / 0.042 / 0.859 \\
\hline
DDIM-SPECT & 28.769 / 0.038 / 0.887 & 28.842 / 0.038 / 0.888 & 28.935 / 0.037 / 0.890 & 28.930 / 0.037 / 0.890 & 28.943 / 0.037 / 0.890  \\
\hline
DiffusionMBIR w/o CT & 27.507 / 0.044 / 0.853 & 28.825 / 0.037 / 0.881 & 29.400 / 0.035 / 0.891 & 29.579 / 0.034 / 0.895 & 29.531 / 0.035 / 0.895 \\
\hline
DiffusionMBIR & 29.655 / 0.038 / 0.895 & 29.741 / 0.034 / 0.906 & 29.832 / 0.034 / 0.907 & 29.819 / 0.034 / 0.907 & 29.826 / 0.034 / 0.908 \\ 
\hline
DDSPECT-3D w/o CT & 27.515 / 0.044 / 0.863 & 28.911 / 0.037 / 0.899 & 29.500 / 0.035 / 0.899 & 29.700 / 0.034 / 0.901 & 30.017 / 0.033 / 0.905 \\
\hline
DDSPECT-3D & 29.945 / 0.036 / 0.901 & 30.008 / 0.034 / 0.905 & 30.054 / 0.033 / 0.913 & 30.162 / 0.033 / 0.914 & 30.331 / 0.032 / 0.915 \\
\hline
\hline
\textbf{DiffSPECT-3D w/o CT}  & 27.954 / 0.042 / 0.867 & 29.577 / 0.035 / 0.893 & 31.434 / 0.028 / 0.916 & 32.570 / 0.025 / 0.927 & 33.324 / 0.023 / 0.934  \\
\hline
\textbf{DiffSPECT-3D (proposed)}  & \textcolor{red}{29.975 / 0.034 / 0.906} & \textcolor{red}{30.078 / 0.033 / 0.909} & \textcolor{red}{31.634 / 0.027 / 0.926} & \textcolor{red}{32.888 / 0.024 / 0.938} & \textcolor{red}{33.702 / 0.022 / 0.946}  \\
\hline
\end{tabular}
}
\label{table0}
\end{table*}

\begin{figure*}[!t]
\centerline{\includegraphics[width=0.8\textwidth]{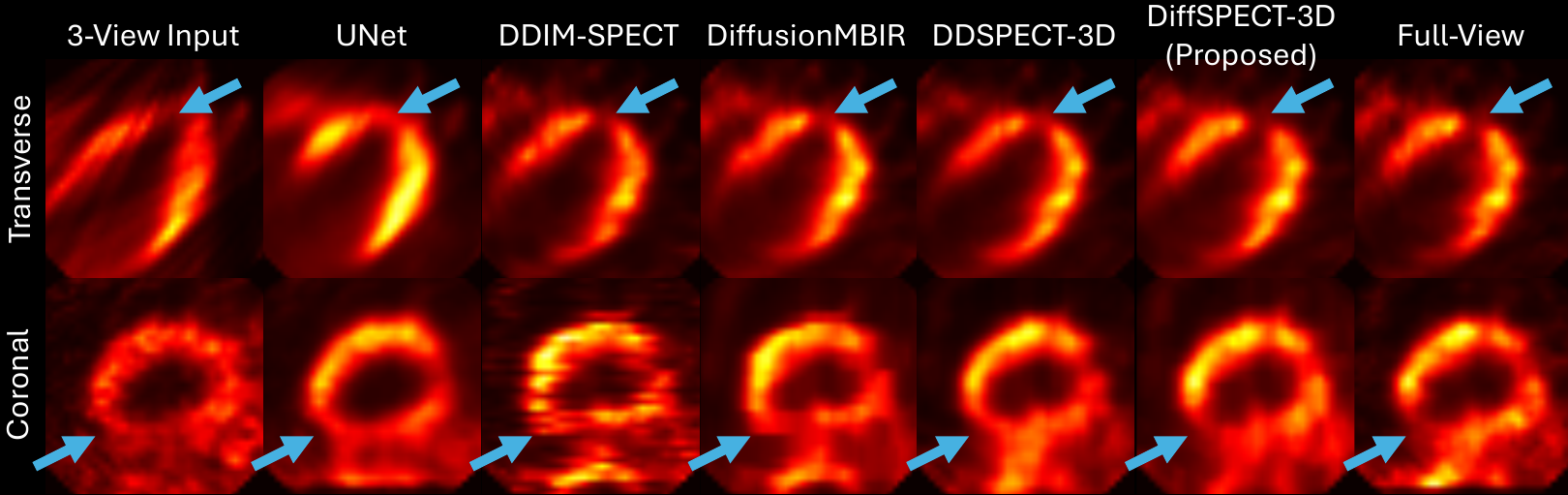}}
\caption{Results from a testing case reconstructed using different methods. Arrows point to the apical perfusion defect of this patient.}
\label{fig_compare}
\end{figure*}

Even though existing deep learning methods cannot be directly applied to SPECT image processing/reconstructions under different acquisition settings due to their limited generalizability, we believe comparisons with other related methods will still be beneficial to show the effectiveness of the proposed method. In this subsection, the proposed method is compared with the following methods:

\begin{enumerate}
\item An U-net-based method proposed for few-view SPECT image post-processing \cite{xie_increasing_2022}. The network follows a U-net-like architecture with squeeze-excitation attention blocks and dense-net blocks inserted after each convolutional layers.
\item The diffusion-based method described in this work \cite{gong_pet_2023} implements 2D DDIM/DDPM sampling for PET image denoising. This study was chosen as it is among the first to apply diffusion models to nuclear image post-processing, demonstrating superior performance compared to U-Net-based models. To ensure a fair comparison with other methods, the DDIM sampling was implemented, and this method is denoted as DDIM-SPECT in this paper.
\item The DiffusionMBIR method introduced in this work \cite{chung_solving_2023} addresses the 3D inconsistency issue by using a 2D diffusion model to mitigate memory burden and improve network performance. This method was chosen because, similar to the proposed DIffSPECT-3D method, the DiffusionMBIR method aims to address the 3D challenges in diffusion model for medical imaging.
\item The dose-aware diffusion model introduced in our previous work \cite{xie_dose-aware_2024} also addresses the 3D challenge in diffusion models. This method was selected due to its superior performance compared to existing models for low-dose/low-count PET image denoising. This work also received the first-place Freek J. Beekman Young Investigator Award at the SNMMI 2024 annual meeting. The network is named as DDPET-3D in the publication \cite{xie_dose-aware_2024} since the paper targets on PET image denoising. To avoid confusion, this method has been re-named as DDSPECT-3D in this paper.
\end{enumerate}
All four methods were trained using both SPECT and CT images as network inputs or as diffusion conditional information. Specifically, the diffusion model becomes $\epsilon_\thetabm{(\xbm_t,t,\xbm_\mathrm{ct}, \xbm_{in})}$. The training data includes SPECT images at five different low-count levels and five different few-view levels. Note that the proposed DiffSPECT-3D does not rely on input SPECT images $\xbm_{in}$ during the training process. Additionally, all the methods were re-trained without the CT images and included for comparison. Quantitative results for the various methods are presented in Table \ref{table0}. One representative testing case was selected and presented in Fig. \ref{fig_compare} to visually compare different methods. The patient study presented in Fig. \ref{fig_compare} is abnormal with a medium sized perfusion defect in the entire apical wall (blue arrows in Fig. \ref{fig_compare}).

As illustrated in Fig. \ref{fig_compare}, the UNet-based method failed to generate optimal results across various acquisition settings, producing over-smoothed images and the perfusion defect was not able to clearly recovered. The DDIM-SPECT, a straightforward 2D diffusion model, resulted in inconsistent images along the coronal view. While DiffusionMBIR addressed the inconsistency issue, it produced images with a distorted myocardium in the coronal view. DDSPECT-3D achieved better recovery of the myocardium contour; however, like the other methods, it smoothed the perfusion defects. The proposed method in this paper, DiffSPECT-3D, demonstrated promising generalizability across various acquisition settings, with better visualization of the perfusion defects.
\section{Discussion and Conclusion}
We introduced DiffSPECT-3D, a diffusion framework for 3D cardiac SPECT imaging that can be generalized to various acquisition conditions. Evaluated on a real-world large-scale clinical cardiac SPECT dataset, with confirmation from cardiac catheterization and diagnostic decisions from nuclear cardiologists, DiffSPECT-3D demonstrates superior quantitative and qualitative results compared to previous baseline methods. DiffSPECT-3D also exhibited superior generalizability when directly applied to data acquired using a different acquisition setting, without further network fine-tuning and re-training. Above certain low-count/few-view levels, DiffSPECT-3D produced images that align with clinical gold-standard, which showed the potential of achieving low-dose/few-view imaging without compromising the diagnostic performance, though a larger-scale validation is necessary to further demonstrate the clinical applicability.

Compared to most of the previous literature \cite{ramon_improving_2020, aghakhan_olia_deep_2022, chen_dudocfnet_2024, du_deep_2024,xie_increasing_2022, xie_deep-learning-based_2023, xie_transformer-based_2023}, which learn the mapping from under-sampled data at one specific acquisition setting to the fully-sampled counterparts, the proposed DiffSPECT-3D can be viewed as a major extension that achieves generalizable reconstructions under different low-count/low-dose and few-view/limited-angle settings and even full-dose/full-view settings. To achieve generalizable reconstructions, the proposed DiffSPECT-3D is featured with \textit{(1)} image-domain consistency using the diffusion posterior sampling strategy \cite{chung2023diffusion}; \textit{(2)} projection-domain consistency by incorporating iterative reconstruction updates with total variation constrain within the diffusion sampling. This ensures that the diffusion prediction at each sampling step aligns with the original under-sampled projection data and the scanner geometry; \textit{(3)} a revised 2.5D conditional strategy that allows the network to observe the entire 3D volume for image reconstructions; \textit{(4)} 3D spatial anatomical information from CT has been incorporated to improve network performance.

The DiffSPECT-3D does not rely on paired under-sampled/fully-sampled SPECT images for network training, and fully-trained network takes the raw projection data as input for image reconstructions. Thus, we believe the DiffSPECT-3D can be easily integrated with current clinical reconstruction workflow to improve SPECT image quality under different acquisition settings.

However, this study has several limitations. First, we were unable to obtain a comprehensive clinical conclusion because physicians’ diagnostic comments and clinical gold standards from cardiac catheterization were not available for all the patient studies included in this work. In the future, with IRB approval, larger-scale validations with clinical gold standards and reader studies involving nuclear cardiologists will be necessary to thoroughly evaluate the clinical performance. We may also quantitatively measured the perfusion defect contrast using artificially created perfusion defects in animal or physical phantom studies. Second, DiffSPECT-3D was only trained and evaluated on $^\text{99m}$Tc tetrofosmin studies acquired on the GE DNM system. Further evaluations on different SPECT tracers and/or different imaging system would be beneficial to extend the algorithm to a broader range of clinical applications.

Despite these limitations, our overall results have been encouraging that DiffSPECT-3D showed the potential of achieving under-sampled imaging without compromising the clinical performance above certain low-count/few-view levels. The experimental results showed that DiffSPECT-3D achieved superior performance compared with previous methods. While this study focuses on SPECT imaging, we believe that the proposed method could be easily extended to other 3D reconstruction tasks for different imaging modalities.

\section*{Acknowledgment}
This work is supported by NIH under grants R01HL154345 and R01EB025468.

\bibliographystyle{ieeetr}
\bibliography{tmi.bib}

\begin{thebibliography}{10}

\bibitem{tsao_heart_2023}
C.~W. Tsao, A.~W. Aday, Z.~I. Almarzooq, C.~A. Anderson, P.~Arora, C.~L. Avery, C.~M. Baker-Smith, A.~Z. Beaton, A.~K. Boehme, A.~E. Buxton, Y.~Commodore-Mensah, M.~S. Elkind, K.~R. Evenson, C.~Eze-Nliam, S.~Fugar, G.~Generoso, D.~G. Heard, S.~Hiremath, J.~E. Ho, R.~Kalani, D.~S. Kazi, D.~Ko, D.~A. Levine, J.~Liu, J.~Ma, J.~W. Magnani, E.~D. Michos, M.~E. Mussolino, S.~D. Navaneethan, N.~I. Parikh, R.~Poudel, M.~Rezk-Hanna, G.~A. Roth, N.~S. Shah, M.-P. St-Onge, E.~L. Thacker, S.~S. Virani, J.~H. Voeks, N.-Y. Wang, N.~D. Wong, S.~S. Wong, K.~Yaffe, S.~S. Martin, and {on behalf of the American Heart Association Council on Epidemiology and Prevention Statistics Committee and Stroke Statistics Subcommittee}, ``Heart {Disease} and {Stroke} {Statistics}—2023 {Update}: {A} {Report} {From} the {American} {Heart} {Association},'' {\em Circulation}, vol.~147, pp.~e93--e621, Feb. 2023.

\bibitem{dobrucki_pet_2010}
L.~W. Dobrucki and A.~J. Sinusas, ``{PET} and {SPECT} in cardiovascular molecular imaging,'' {\em Nature Reviews Cardiology}, vol.~7, pp.~38--47, Jan. 2010.

\bibitem{abbott_contemporary_2018}
B.~G. Abbott, J.~A. Case, S.~Dorbala, A.~J. Einstein, J.~R. Galt, R.~Pagnanelli, R.~P. Bullock-Palmer, P.~Soman, and R.~G. Wells, ``Contemporary {Cardiac} {SPECT} {Imaging}—{Innovations} and {Best} {Practices}: {An} {Information} {Statement} from the {American} {Society} of {Nuclear} {Cardiology},'' {\em Journal of Nuclear Cardiology}, vol.~25, pp.~1847--1860, Oct. 2018.

\bibitem{lin_radiation_2010}
E.~C. Lin, ``Radiation {Risk} {From} {Medical} {Imaging},'' {\em Mayo Clinic Proceedings}, vol.~85, pp.~1142--1146, Dec. 2010.

\bibitem{wang_deep_2020}
G.~Wang, J.~C. Ye, and B.~De~Man, ``Deep learning for tomographic image reconstruction,'' {\em Nature Machine Intelligence}, vol.~2, pp.~737--748, Dec. 2020.

\bibitem{ramon_improving_2020}
A.~J. Ramon, Y.~Yang, P.~H. Pretorius, K.~L. Johnson, M.~A. King, and M.~N. Wernick, ``Improving {Diagnostic} {Accuracy} in {Low}-{Dose} {SPECT} {Myocardial} {Perfusion} {Imaging} {With} {Convolutional} {Denoising} {Networks},'' {\em IEEE Transactions on Medical Imaging}, vol.~39, pp.~2893--2903, Sept. 2020.
\newblock Conference Name: IEEE Transactions on Medical Imaging.

\bibitem{aghakhan_olia_deep_2022}
N.~Aghakhan~Olia, A.~Kamali-Asl, S.~Hariri~Tabrizi, P.~Geramifar, P.~Sheikhzadeh, S.~Farzanefar, H.~Arabi, and H.~Zaidi, ``Deep learning–based denoising of low-dose {SPECT} myocardial perfusion images: quantitative assessment and clinical performance,'' {\em Eur J Nucl Med Mol Imaging}, vol.~49, pp.~1508--1522, Apr. 2022.

\bibitem{chen_dudocfnet_2024}
X.~Chen, B.~Zhou, X.~Guo, H.~Xie, Q.~Liu, J.~S. Duncan, A.~J. Sinusas, and C.~Liu, ``{DuDoCFNet}: {Dual}-{Domain} {Coarse}-to-{Fine} {Progressive} {Network} for {Simultaneous} {Denoising}, {Limited}-{View} {Reconstruction}, and {Attenuation} {Correction} of {Cardiac} {SPECT},'' {\em IEEE Transactions on Medical Imaging}, vol.~43, pp.~3110--3125, Sept. 2024.

\bibitem{du_deep_2024}
Y.~Du, J.~Sun, C.-Y. Li, B.-H. Yang, T.-H. Wu, and G.~S.~P. Mok, ``Deep learning-based multi-frequency denoising for myocardial perfusion {SPECT},'' {\em EJNMMI Physics}, vol.~11, p.~80, Oct. 2024.

\bibitem{xie_increasing_2022}
H.~Xie, S.~Thorn, X.~Chen, B.~Zhou, H.~Liu, Z.~Liu, S.~Lee, G.~Wang, Y.-H. Liu, A.~J. Sinusas, and C.~Liu, ``Increasing angular sampling through deep learning for stationary cardiac {SPECT} image reconstruction,'' {\em J. Nucl. Cardiol.}, May 2022, DOI: 10.1007/s12350-022-02972-z.

\bibitem{xie_deep-learning-based_2023}
H.~Xie, S.~Thorn, Y.-H. Liu, S.~Lee, Z.~Liu, G.~Wang, A.~J. Sinusas, and C.~Liu, ``Deep-{Learning}-{Based} {Few}-{Angle} {Cardiac} {SPECT} {Reconstruction} {Using} {Transformer},'' {\em IEEE Transactions on Radiation and Plasma Medical Sciences}, vol.~7, pp.~33--40, Jan. 2023.
\newblock Conference Name: IEEE Transactions on Radiation and Plasma Medical Sciences.

\bibitem{xie_transformer-based_2023}
H.~Xie, B.~Zhou, X.~Chen, X.~Guo, S.~Thorn, Y.-H. Liu, G.~Wang, A.~Sinusas, and C.~Liu, ``Transformer-{Based} {Dual}-{Domain} {Network} for {Few}-{View} {Dedicated} {Cardiac} {SPECT} {Image} {Reconstructions},'' in {\em Medical {Image} {Computing} and {Computer} {Assisted} {Intervention} – {MICCAI} 2023}, (Cham), pp.~163--172, Springer Nature Switzerland, 2023.

\bibitem{xie_increasing_2024}
H.~Xie, A.~Alashi, S.~Thorn, X.~Chen, B.~Zhou, A.~Sinusas, and C.~Liu, ``Increasing angular sampling for dedicated cardiac {SPECT} imaging guided by deep learning: {Validation} with human data,'' {\em Journal of Nuclear Medicine}, vol.~65, pp.~241570--241570, June 2024.

\bibitem{Croitoru.etal2023}
F.-A. Croitoru, V.~Hondru, R.~T. Ionescu, and M.~Shah, ``Diffusion models in vision: A survey,'' {\em IEEE Transactions on Pattern Analysis and Machine Intelligence}, 2023.

\bibitem{Kazerouni.etal2023a}
A.~Kazerouni, E.~K. Aghdam, M.~Heidari, R.~Azad, M.~Fayyaz, I.~Hacihaliloglu, and D.~Merhof, ``Diffusion models for medical image analysis: A comprehensive survey,'' {\em arXiv preprint arXiv:2211.07804}, 2022.

\bibitem{ho2020denoising}
J.~Ho, A.~Jain, and P.~Abbeel, ``Denoising diffusion probabilistic models,'' {\em arXiv preprint arxiv:2006.11239}, 2020.

\bibitem{gong_pet_2023}
K.~Gong, K.~Johnson, G.~El~Fakhri, Q.~Li, and T.~Pan, ``{PET} image denoising based on denoising diffusion probabilistic model,'' {\em European Journal of Nuclear Medicine and Molecular Imaging}, 2023.

\bibitem{xie_dose-aware_2024}
H.~Xie, W.~Gan, B.~Zhou, M.-K. Chen, M.~Kulon, A.~Boustani, B.~A. Spencer, R.~Bayerlein, W.~Ji, X.~Chen, Q.~Liu, X.~Guo, M.~Xia, Y.~Zhou, H.~Liu, L.~Guo, H.~An, U.~S. Kamilov, H.~Wang, B.~Li, A.~Rominger, K.~Shi, G.~Wang, R.~D. Badawi, and C.~Liu, ``Dose-aware {Diffusion} {Model} for {3D} {Low}-dose {PET}: {Multi}-institutional {Validation} with {Reader} {Study} and {Real} {Low}-dose {Data},'' Sept. 2024.
\newblock arXiv:2405.12996.

\bibitem{gao_corediff_2023}
Q.~Gao, Z.~Li, J.~Zhang, Y.~Zhang, and H.~Shan, ``{CoreDiff}: Contextual error-modulated generalized diffusion model for low-dose {CT} denoising and generalization,'' {\em {IEEE} Transactions on Medical Imaging}, pp.~1--1, 2023.

\bibitem{gungor_adaptive_2023}
A.~Güngör, S.~U. Dar, Å.~Öztürk, Y.~Korkmaz, H.~A. Bedel, G.~Elmas, M.~Ozbey, and T.~Çukur, ``Adaptive diffusion priors for accelerated {MRI} reconstruction,'' {\em Medical Image Analysis}, vol.~88, p.~102872, 2023.

\bibitem{chung_score-based_2022}
H.~Chung and J.~C. Ye, ``Score-based diffusion models for accelerated {MRI},'' {\em Medical Image Analysis}, vol.~80, p.~102479, 2022.

\bibitem{chung_solving_2023}
H.~Chung, D.~Ryu, M.~T. McCann, M.~L. Klasky, and J.~C. Ye, ``Solving 3d inverse problems using pre-trained 2d diffusion models,'' in {\em Proceedings of the IEEE/CVF Conference on Computer Vision and Pattern Recognition (CVPR)}, pp.~22542--22551, June 2023.

\bibitem{lee_improving_2023}
S.~Lee, H.~Chung, M.~Park, J.~Park, W.-S. Ryu, and J.~C. Ye, ``Improving 3d imaging with pre-trained perpendicular 2d diffusion models,'' 2023.

\bibitem{ronneberger_u-net_2015}
O.~Ronneberger, P.~Fischer, and T.~Brox, ``U-{Net}: {Convolutional} {Networks} for {Biomedical} {Image} {Segmentation},'' in {\em Medical {Image} {Computing} and {Computer}-{Assisted} {Intervention} – {MICCAI} 2015}, pp.~234--241, Springer International Publishing, 2015.

\bibitem{xie_unified_2023}
H.~Xie, Q.~Liu, B.~Zhou, X.~Chen, X.~Guo, H.~Wang, B.~Li, A.~Rominger, K.~Shi, and C.~Liu, ``Unified {Noise}-{Aware} {Network} for {Low}-{Count} {PET} {Denoising} {With} {Varying} {Count} {Levels},'' {\em IEEE Transactions on Radiation and Plasma Medical Sciences}, 2023.

\bibitem{xie_segmentation-free_2023}
H.~Xie, Z.~Liu, L.~Shi, K.~Greco, X.~Chen, B.~Zhou, A.~Feher, J.~C. Stendahl, N.~Boutagy, T.~C. Kyriakides, G.~Wang, A.~J. Sinusas, and C.~Liu, ``Segmentation-free {PVC} for cardiac {SPECT} using a densely-connected multi-dimensional dynamic network,'' {\em {IEEE} Transactions on Medical Imaging}, vol.~42, no.~5, pp.~1325--1336, 2023.

\bibitem{li2022odconv}
C.~Li, A.~Zhou, and A.~Yao, ``Omni-dimensional dynamic convolution,'' in {\em International Conference on Learning Representations}, 2022.

\bibitem{xie_noise-aware_2025}
H.~Xie, L.~Guo, A.~Velo, Z.~Liu, Q.~Liu, X.~Guo, B.~Zhou, X.~Chen, Y.-J. Tsai, T.~Miao, M.~Xia, Y.-H. Liu, I.~S. Armstrong, G.~Wang, R.~E. Carson, A.~J. Sinusas, and C.~Liu, ``Noise-aware dynamic image denoising and positron range correction for {Rubidium}-82 cardiac {PET} imaging via self-supervision,'' {\em Medical Image Analysis}, vol.~100, p.~103391, 2025.

\bibitem{chung2023diffusion}
H.~Chung, J.~Kim, M.~T. Mccann, M.~L. Klasky, and J.~C. Ye, ``Diffusion posterior sampling for general noisy inverse problems,'' in {\em The Eleventh International Conference on Learning Representations}, 2023.

\bibitem{bocher_fast_2010}
M.~Bocher, I.~M. Blevis, L.~Tsukerman, Y.~Shrem, G.~Kovalski, and L.~Volokh, ``A fast cardiac gamma camera with dynamic {SPECT} capabilities: design, system validation and future potential,'' {\em European Journal of Nuclear Medicine and Molecular Imaging}, vol.~37, pp.~1887--1902, Oct. 2010.

\bibitem{chan_impact_2016}
C.~Chan, J.~Dey, Y.~Grobshtein, J.~Wu, Y.-H. Liu, R.~Lampert, A.~J. Sinusas, and C.~Liu, ``The impact of system matrix dimension on small {FOV} {SPECT} reconstruction with truncated projections,'' {\em Medical Physics}, vol.~43, no.~1, pp.~213--224, 2016.

\bibitem{shepp_maximum_1982}
L.~Shepp and Y.~Vardi, ``Maximum {Likelihood} {Reconstruction} for {Emission} {Tomography},'' {\em IEEE Transactions on Medical Imaging}, vol.~1, pp.~113--122, Oct. 1982.

\bibitem{lange_em_1984}
K.~Lange and R.~Carson, ``{EM} reconstruction algorithms for emission and transmission tomography,'' {\em Journal of Computer Assisted Tomography}, vol.~8, pp.~306--316, Apr. 1984.

\bibitem{dhariwal2021diffusion}
P.~Dhariwal and A.~Nichol, ``Diffusion models beat gans on image synthesis,'' {\em Advances in neural information processing systems}, vol.~34, pp.~8780--8794, 2021.

\bibitem{nichol2021improved}
A.~Q. Nichol and P.~Dhariwal, ``Improved denoising diffusion probabilistic models,'' in {\em International Conference on Machine Learning}, pp.~8162--8171, PMLR, 2021.

\bibitem{boyd_distributed_2011}
S.~Boyd, N.~Parikh, E.~Chu, B.~Peleato, and J.~Eckstein, ``Distributed {Optimization} and {Statistical} {Learning} via the {Alternating} {Direction} {Method} of {Multipliers},'' {\em Foundations and Trends® in Machine Learning}, vol.~3, pp.~1--122, July 2011.
\newblock Publisher: Now Publishers, Inc.

\bibitem{persson_total_2001}
M.~Persson, D.~Bone, and H.~Elmqvist, ``Total variation norm for three-dimensional iterative reconstruction in limited view angle tomography,'' {\em Physics in Medicine \& Biology}, vol.~46, p.~853, Mar. 2001.

\bibitem{song_denoising_2022}
J.~Song, C.~Meng, and S.~Ermon, ``Denoising diffusion implicit models,'' {\em arXiv preprint arXiv:2010.02502}, 2022.

\bibitem{dvorak_interpretation_2011}
R.~A. Dvorak, R.~K.~J. Brown, and J.~R. Corbett, ``Interpretation of {SPECT}/{CT} {Myocardial} {Perfusion} {Images}: {Common} {Artifacts} and {Quality} {Control} {Techniques},'' {\em RadioGraphics}, vol.~31, pp.~2041--2057, Nov. 2011.

\end{thebibliography}

\end{document}